\documentclass[prl,twocolumn,showpacs,preprintnumbers,amsmath,amssymb]{revtex4}

\usepackage{graphicx}
\usepackage{dcolumn}
\usepackage{bm}
\usepackage{amsmath}
\usepackage{epsfig}
\usepackage{ae}

\begin{document}

\preprint{June 2003}

\title{Successive Umbrella Sampling}

\author{Peter Virnau}
 \email{peter.virnau@uni-mainz.de}
\author{Marcus M\"uller}
\affiliation{%
Institut f\"ur Physik, WA331, Universit\"at Mainz, Staudinger Weg 7, 55099 Mainz, Germany
}%

\date{\today}

\begin{abstract}
We propose an extension of umbrella sampling in which the pertinent range of
states is subdivided in windows that are sampled consecutively and linked
together. Extrapolating results from one window we estimate a weight function
for the neighboring window.  We present a detailed analysis and demonstrate
that the error is controlled and independent from the window sizes. The analysis also
allows us to detect sampling difficulties.  The efficiency of the algorithm is
comparable to a multicanonical simulation with an ideal weight function. We
exemplify the computational scheme by simulating the liquid-vapor coexistence
in a Lennard--Jones system.
%
%
\end{abstract}
\pacs{02.70.Tt, 05.10.Ln, 05.70.Fh, 64.70.Fx}
\maketitle
%
%
%
%

Calculating and overcoming free energy barriers remains a considerable challenge in computational 
condensed and soft matter physics. Applications are manyfold and range from 
protein folding \cite{Hansmann} over quantum systems \cite{TROYER} to the study of first 
order phase transitions \cite{Gibbs}, nucleation or glassy systems \cite{MARINARI}.

Various sophisticated sampling techniques have been devised to estimate free
energy differences.  Though methods are general we shall use the language of a
liquid-vapor transition, where the number of particles $n$ is the order
parameter of the transition. In this case the Helmholtz free energy $A$ can be
computed from the probability distribution $P[n]$: $A[n]=-k_{\rm B}T\ln P[n]+{\rm const.}$,
where $k_{\rm B}$ stands for Boltzmann's constant and $T$ for temperature. Free energy
barriers correspond to regions of extremely low probability, which make efficient 
sampling difficult.

Multicanonical methods \cite{multican1} modify the Hamiltonian in order
to sample a range of densities uniformly. To this end, one adds 
a weight function $w[n]$ to the original Hamiltonian such that the 
simulated distribution $P_{\rm sim}[n]=P[n]\exp(-w[n])$ becomes flat 
for the choice of $w[n]\approx\ln P[n]$. Unfortunately, $P[n]$ is {\em a
priori} unknown, and several methods have been explored to estimate $w[n]$:
(i) Histogram-reweighting techniques \cite{histo_rew} alleviate this problem by performing a 
sequence of weighted simulations and extrapolations starting at a point where barriers 
are small and the system explores a broad range of $n$.
More sophisticated methods combine results of multiple histograms \cite{Swendsen}.
(ii) Multicanonical Recursion \cite{multiRec} conducts a series of short trial
runs. After each run $w[n]$ is adjusted until the simulation can access all
relevant states. The weight factors can also be self-adjusted during the
simulation \cite{WangLandau,dePablo,Hansmann,jesko}. However, detailed balance \cite{Frenkel} is
violated in this process and separation of statistical and systematic errors
becomes difficult.
(iii) Weight factors can also be obtained from the transition probabilities
between macrostates \cite{TMMC, ADB,Shell}.

In umbrella sampling \cite{umbrella} the pertinent
range of macrostates, defined by the number of particles $n$, is subdivided
into $m$ overlapping windows of length $\omega$.  In this work we propose an
extension in which adjacent windows are sampled consecutively. Successive
umbrella sampling offers substantial advantages: It allows us to generate weight factors
for subsequent windows by means of extrapolation.  In contrast to histogram-reweighting,
simulations can be performed anywhere in the phase diagram without prior
knowledge of a weight function.  In contrast to self-adjusting methods
\cite{WangLandau,dePablo,Hansmann,jesko}, $w[n]$ stays fixed during the run and
detailed balance is not violated.  Although these powerful schemes are suitable
for generating weight functions, they require in principle an additional
multicanonical production run whenever errors need to be determined exactly.
In our approach, every run contributes uniformly to statistics.  Additionally,
it does not involve any adjustable parameters (e.g., the modification factor
$f$ in Ref.~\cite{WangLandau}).  

\begin{figure}[htp]
\begin{center}
\epsfig{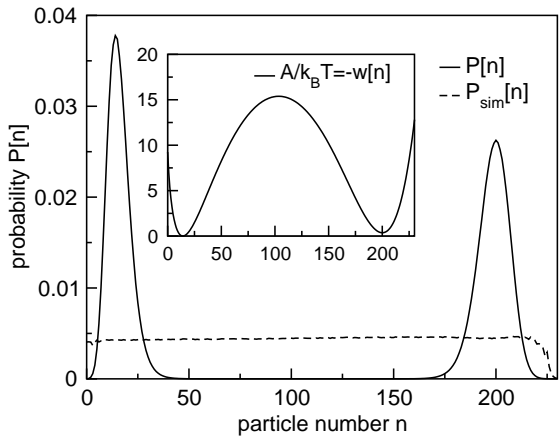}
\caption{P[n] for a LJ system at T=0.85437~$\frac{\epsilon}{k}=0.85~T_{\rm c}$, 
$L=6.74~\sigma$. The system is at coexistence because the area below both
peaks is equal \cite{EWR}. Else, coexistence can be established by:
P$_{\rm coex}[n]=P[n]\cdot\exp((\mu_{\rm coex}-\mu)nk_{\rm B}T)$. 
Dashed lines: $P_{\rm sim}[n]$ from a multicanonical simulation with a very
good weight function $w[n]$.
Inset: free energy barrier.}
\label{fig:prob}
\end{center}
\end{figure}

%
%
%
In the following sections we demonstrate
the method by simulating the liquid-vapor coexistence in the $\mu VT$-ensemble.
Beads interact via a truncated and shifted Lennard--Jones 
potential with parameters $\epsilon$ and $\sigma$:
\begin{equation}
V_\text{LJ}=\begin{cases}4\epsilon\cdot\left[\left(\frac{\sigma}{r}\right)^{12}-\left(\frac{\sigma}{r}\right)^{6}+\frac{127}{16384}\right], & \text{if} \hspace{0.2cm} r\leq 2\cdot 2^{1/6}\sigma , \\
0, & \text{else}.
\end{cases}
\label{LJ_pot}
\end{equation} 
The box length of our reference system is $6.74~\sigma$,
temperature $T=0.85437~\frac{\epsilon}{k}$.
Fig.~\ref{fig:prob} shows the probability distribution
as a function of particle number $n$ at coexistence.
The peak at low density corresponds to the gas phase, the peak at high density 
to the liquid phase. 
Both are separated by a free energy barrier (here: $15.2~k_{\rm B}T$).
\noindent

%
%
%
%
The guiding idea of successive umbrella sampling is to investigate one small
window after the other, starting at zero density. A histogram $H_k[n]$ monitors how often each state is
visited in the $k$th window $[k\omega,(k+1)\omega]$.  Let $H_{kl}\equiv H_k[k\omega]$ and
$H_{kr}\equiv H_k[(k+1)\omega]$ denote the values of the $k$th histogram at its left
and right boundary, respectively, and $r_k \equiv H_{kr}/H_{kl}$ characterize their
ratio. After a predetermined number of insertion/deletion Monte Carlo (MC) moves per window, 
the (unnormalized) probability distribution can be estimated recursively:
\begin{equation}
\frac{P[n]}{P[0]}=\frac{H_{0r}}{H_{0l}}\cdot\frac{H_{1r}}{H_{1l}}\cdots\frac{H_k[n]}{H_{kl}}
                 = \Pi_{i=1}^{k-1}r_i \cdot \frac{H_k[n]}{H_{kl}}
\label{prob_eq}
\end{equation}
when $n \in [k\omega,(k+1)\omega]$. Probability ratios in Eq.~(\ref{prob_eq}) correspond to free energy differences. 
Care has to be taken at the boundaries of a window to fulfill detailed balance 
\cite{footnote}. 

As we are sampling one window after the other, efficiency can be increased by 
combining the algorithm with the multicanonical concept.
In a weighted simulation we replace $H[n]$ in Eq.~\eqref{prob_eq} by $H[n]\exp(w[n])$. 
A good estimate for the weight function may be obtained by extrapolation: After
$P[n]$ is determined according to Eq.~\eqref{prob_eq}, $w[n]=\ln(P[n])$ is
extrapolated quadratically into the next window. 
The first window is usually unweighted. If in this case states are not accessible,
$w[n]$ might be altered by a
fixed number of $k_{\rm B}T$ in each iteration step.  

\begin{figure}[thb]
\begin{center}
\epsfig{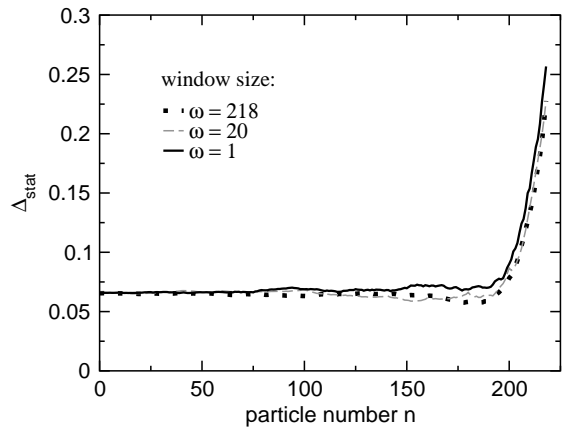}
\caption{Comparison of (normalized) weighted umbrella sampling runs with 
different window sizes and an average number of 2 million MC steps per state \cite{footnote}: 
(a) one section runs with window size $\omega=218$, (b) $m=11$ section runs with 
window size $\omega=20$, (c) $\omega=1$ (i.e., two states per window).
The precision of the factors $\exp (w(n))$, which have been used to generate the data, is better than $2\%$ for (a) and (b) and $30\%$ for (c).
Errors correspond to single runs and were determined as the
standard deviation over 400 runs.}
\label{fig:mc}
\end{center}
\end{figure}
\begin{figure}[thb]
\begin{center}
\epsfig{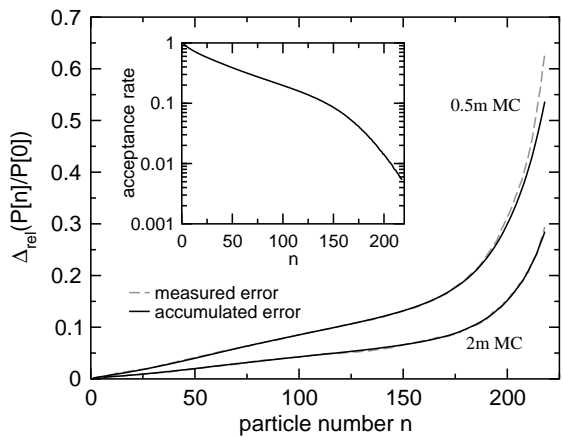}
\caption{Comparison of accumulated statistical error (Eq.~\eqref{stat_tot}) and independently 
measured error for ($\omega$=1):
(a) 4000 runs with 0.5 million MC steps per state \protect\cite{footnote}, 
(b) 400 runs with 2 million MC steps per state. Measured and accumulated error cannot be distinguished.
Inset: acceptance rate of insertion/deletion attempts. 
}
\label{fig:corr}
\end{center}
\end{figure}

Let us investigate how the computational effort depends on the choice 
of the window size. It has been suggested in the literature \cite{Chandler,Frenkel}
that small windows reduce computational effort by a factor of $m$: if $\tau$ 
is the time to obtain a predetermined statistical error $\Delta_1$ in 
a single window, then $\tau\propto\omega^{2}/\Delta_1$ and computation time
$t_{\rm cpu}\propto m\omega^2/\Delta_1$. In the limit of a single large window, m'=1, this yields
$t_{\rm cpu}'=(m\omega)^{2}=mt_{\rm cpu}$. In Fig.~\ref{fig:mc} we compare the error 
which was obtained from multicanonical runs with different window sizes but an equal total number of Monte Carlo steps \cite{footnote}.
All distributions were normalized to unity before the error
was calculated. As expected, errors are evenly distributed for low and intermediate densities.
For large densities deviations occur and the relative error becomes larger. 
The absolute value of $P[n]$, however, is small in this region (cf.\ Fig.~\ref{fig:prob})
such that the error does not influence normalization. Contrary to the
simple argument above the error is roughly independent of the window size
$\omega$. To simplify the discussion, one
can also set $P[0] \equiv 1$ and let errors accumulate from low to high densities (cf.\ Fig.~\ref{fig:corr}). 
Using Eq.~\eqref{prob_eq} and assuming 
that there are no correlations between neighboring intervals we obtain
\begin{equation}
\Delta\left( \frac{P[n]}{P[0]}\right) =\sqrt{\sum_{i=0}^{k-1}\Delta r_i^{2} + \Delta \left(\frac{H_k[n]}{H_{kl}}\right)^{2}} \sim {\cal O}(\Delta_1\sqrt{k}).
\label{stat_tot}
\end{equation}
If errors are comparable for all windows, the overall error for $m$ windows is of order
{\cal O}($\sqrt{m}$) bigger as for $m=1$. 
To compensate this, sampling time in each window
has to be increased by a factor of $m$.  Hence, computational effort for a given
error is always of order $(m\omega)^{2}=N^{2}$ and does not depend on the
number of windows into which the sampling range is subdivided. 
We conclude that successive umbrella sampling is as fast as a 
multicanonical simulation with a very good weight function.
Note further that a rather inaccurate estimate for $w[n]$ leads only to a
slight increase in the error (see figure caption in Fig.~\ref{fig:mc}).
In the following we focus  on the smallest window size $\omega=1$.
One should keep in mind, however, that weighting in a single variable might not always be 
sufficient when free energy landscapes become more complex \cite{EX}.

Subsequently, we present a detailed error analysis of our method. 
Eq.~\eqref{stat_tot} only accounts for statistical errors within an interval and neglects 
correlations between adjacent windows and possible systematic errors when estimating probability ratios.
The former may occur, because we use the ending 
configuration of the $k$th window as the starting configuration of the $(k+1)$th 
window. Accumulating the relative statistical errors of individual ratios $\Delta r_i$ in Eq.~\eqref{stat_tot}, 
and comparing the result with the independently measured statistical error, 
we can gauge the importance of correlations between neighboring windows.
In Fig.~\ref{fig:corr} two features can be
identified.  First, large errors are obtained for high densities because it becomes
more and more difficult to insert particles (cf.\ Fig.~\ref{fig:corr} inset). 
Secondly, we find good agreement between both curves if the number 
of Monte Carlo steps per state is large. If the computational effort is reduced, curves only match for 
low and medium densities because in these regions correlations decay fast.
Although this test is not crucial as 
errors can be measured directly, it enables us to check if high 
density configurations are well equilibrated. 
If one is only interested in a rough estimate, the total error of a simulation can be
preset before sampling starts. For this purpose,
the termination condition is changed from a given number of Monte Carlo steps to a predefined error $\Delta_1$
in each window. Then, the total error adds up to $\sqrt{m}\cdot \Delta_1$ according to Eq.~\eqref{stat_tot}
\cite{footnote2}.

\begin{figure}[htp]
\begin{center}
\epsfig{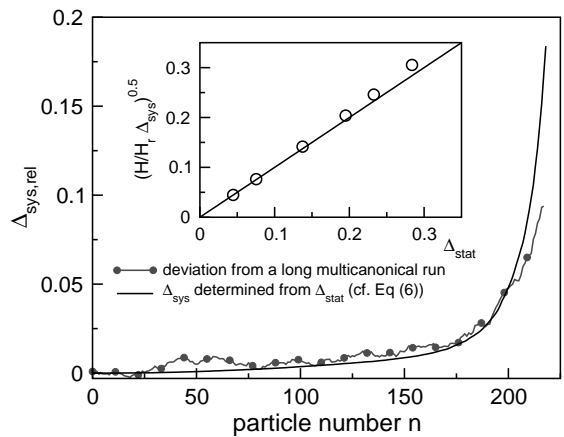}
\caption{
Relative systematic errors for a run with 0.5 million MC steps per state (compare with Fig.~\eqref{fig:corr}).
Inset: Relation between relative systematic and statistical error. Line 
corresponds to Eq.~\eqref{Delta_eq}, $\circ$ denotes the relations determined by simulation. 
}
\label{fig:error}
\end{center}
\end{figure}
In addition to the statistical error, the use of Eq.~\eqref{prob_eq} imparts a
small systematic error onto the probability distribution because the average of a probability 
ratio is larger than the ratio over two probability averages. An expansion of the former leads to
\begin{equation}
\langle r_i \rangle \equiv \left\langle \frac{H_{ir}}{H_{il}}\right\rangle
= \frac{\langle H_{ir}\rangle}{\langle H_{il} \rangle} 
 \left( 1 +\frac{\langle \delta H_{il}^2\rangle}{\langle H_{il}\rangle^2}
                   -\frac{\langle \delta H_{il} \delta H_{ir}\rangle}{\langle H_{il}\rangle\langle H_{ir}\rangle}
\cdots \right)
\label{H_eq}
\end{equation}
where $\delta H_{il} \equiv H_{il} - \langle H_{il} \rangle$ and similarly 
$\delta H_{ir}\equiv H_{ir} - \langle H_{ir}\rangle$.
$\langle H_{ir}/H_{ir}\rangle$ corresponds to the 
probability ratio as it is measured in simulation whereas 
$\langle H_{ir} \rangle/\langle H_{ir}\rangle$  is equal to $P$[(i+1)$\omega$]/$P$[i$\omega$]. 
It is therefore advantageous to determine the probability ratio as the ratio of the sums 
of $H_{ir}$ and $H_{il}$ over all runs 
(and not as the sums over the ratios). 
The last factor is typically larger than unity because fluctuations 
of $H_{il}$ and $H_{ir}$ are anticorrelated. 

For the special case $\omega=1$, $H_{il}+H_{ir}=H$ where $H$ denotes the
total number of entries in the $i$th histogram. In this case, the last factor in Eq.~(\ref{H_eq}) takes the form
$1+\Delta_{\rm sys}^{\omega=1}r_i$ with a relative systematic error
\begin{equation}
\Delta_{\rm sys}^{\omega=1}r_i = \frac{H}{\langle H_{il}\rangle \langle H_{ir}\rangle} \frac{\langle \delta H_{il}^2\rangle}{\langle H_{il}\rangle}.
\label{eq:sys_err}
\end{equation}
In the context of a grandcanonical simulation, the systematic error corresponds to a shift of the
chemical potential of the order $k_BT\Delta_{\rm sys}^{\omega=1}r_i$. Comparing this value with
the relative statistical error $\Delta r_i$ of the ratio, we obtain
\begin{equation}
\Delta_{\rm sys}^{\omega=1}r_i \approx \Delta r_i^2 \frac{\langle H_{ir}\rangle}{H}.
\label{Delta_eq}
\end{equation}
Relation \eqref{Delta_eq} allows us to compute the systematic error from the known statistical error. 
To test the approximation we determined an accurate
weight function for $P$(n=14)/$P$(n=13) (gas peak) by a long run. $\langle$$H$(14)/$H$(13)$\rangle$
was computed by averaging over a large number of short runs with $H$=30, 50, 80, 200, 800 and 2400 MC 
steps, respectively. The true
systematic offset is the deviation of this ratio from 1. 
From the results in the inset of Fig.~\ref{fig:error} we conclude that
Eqs.~(\ref{eq:sys_err}, \ref{Delta_eq}) yield an excellent approximation for the systematic error, 
even when errors become large. For $\omega>1$ we expect that the systematic error of a single ratio 
is also of the order of $\Delta r_i^2$.

In analogy to Eq.~\eqref{stat_tot} we obtain the total relative systematic error of $P[n]/P[0]$:
\begin{eqnarray}
\Delta_{\rm sys}&&\left( \frac{P[n]}{P[0]}\right) = \sum_{i=1}^{k-1}\Delta_{\rm sys}r_i + \Delta_{\rm sys} \left( \frac{H_k[n]}{H_{kl}}\right)  \\
                &&                        \sim {\cal O}(\Delta_{\rm sys,1}k) \sim {\cal O}(\Delta_1^2k) \sim {\cal O}\left(\Delta\left( \frac{P[n]}{P[0]}\right)\right)^2.
                                                \nonumber
\label{sys_tot}
\end{eqnarray}
where $\Delta_{\rm sys,1}$ denotes the systematic relative error of a single ratio. 
Both, the total and the individual systematic errors scale like the square of the statistical errors. Hence, if the
statistical error is small the systematical error is negligible.  By the same token, however, the method is not suitable 
for quickly generating a rough estimate of weight factors (to be used in a subsequent multicanonical simulation).
If the statistical error becomes comparable to unity the systematic error will be of the same order.

In Fig.~\eqref{fig:error} we analyze the systematic error propagation for the runs with $\omega=1$ and 500,000 MC steps 
per state, which exhibit rather large statistical errors (cf.\ Fig.~3).
Firstly, we regard the deviation between the probability distribution obtained from those short runs and an estimate for
the true distribution obtained from a very long multicanonical run over the entire range. At small particle numbers the systematic 
deviation agrees with the values calculated from Eq.~\eqref{eq:sys_err} within the accuracy of the true estimate ($0.5\%-2\%$).
At large particle number, however, the calculated value is larger than the deviation from the true distribution. In this region,
we systematically underestimate the ratios $r_i$ upon increasing the number of particles, because we fail 
to equilibrate the configurations on the time scale of the short simulation runs. Therefore configurations with lower
particle number (contributing to $H_{il}$) are favored. These correlation effects are not included
in Eq.~\eqref{eq:sys_err}. Since both systematical errors have opposite signs Eq.~\eqref{eq:sys_err} constitutes an upper bound.

In summary, we propose an extension to umbrella sampling in which simulation
windows are sampled successively. This allows us to generate weight factors 
by extrapolating results from previous windows.  Hence, the scheme is able to calculate free energy
barriers without prior knowledge of weight factors.  The computational
efficiency is competitive to alternative approaches which adjust
weight factors on the fly and in principle independent from the choice of the
window size.  The algorithm is straightforward to implement and to parallelize.
It involves neither fine-tuning of simulation parameters, nor violation of
detailed balance.  Errors can be calculated exactly and are controlled.  A
detailed analysis revealed the existence of a small systematic error which is,
however, irrelevant in practice.  We also provided a scheme to check if configurations are
well equilibrated.  Our algorithm can be readily applied to a large variety of
problems which involve the computation of free energy barriers.  Possible 
applications include protein folding, the study
of interfaces, nucleation barriers, and first order phase transitions. 

It is a great pleasure to thank K. Binder and L.G. MacDowell for stimulating discussions.
Financial support by BASF AG and DFG under grant Bi314/17 is gratefully 
acknowledged. CPU time was provided by the NIC J\"ulich and the HLRZ Stuttgart. 

\bibliography{ref}

\end{document}